\documentclass[epj]{webofc}
\usepackage[varg]{txfonts}   
\woctitle{MESON2014 - the 13$^\textrm{th}$ International Workshop on 
Meson Production, Properties and Interaction}
\begin{document}
\selectlanguage{english}
\title{In-medium $\bar K$ interactions and bound states} 


\author{Avraham Gal\inst{1}\fnsep\thanks{\email{avragal@savion.huji.ac.il}} 
\and Eli Friedman\inst{1} \and Nir Barnea\inst{1} \and 
Ale\v{s} Ciepl\'{y}\inst{2} \and Ji\v{r}\'{i} Mare\v{s}\inst{2} \and 
Daniel Gazda\inst{3}} 

\institute{Racah Institute of Physics, The Hebrew University, Jerusalem 91904, 
Israel \and Nuclear Physics Institute, 25068 \v{R}e\v{z}, Czech Republic 
\and ECT*, Villa Tambosi, I-38123 Villazzano (Trento), Italy} 

\abstract{Correct treatment of subthreshold $\bar K N$ dynamics is mandatory 
in $K^-$-atom and $\bar K$-nuclear bound-state calculations, as demonstrated 
by using in-medium chirally-based models of $\bar K N$ interactions. Recent 
studies of kaonic atom data reveal appreciable multi-nucleon contributions. 
$\bar K$-nuclear widths larger than 50 MeV are anticipated.} 
\maketitle
\section{Introduction}
\label{intro}

\begin{figure}[htb] 
\centering 
\includegraphics[width=0.9\textwidth]{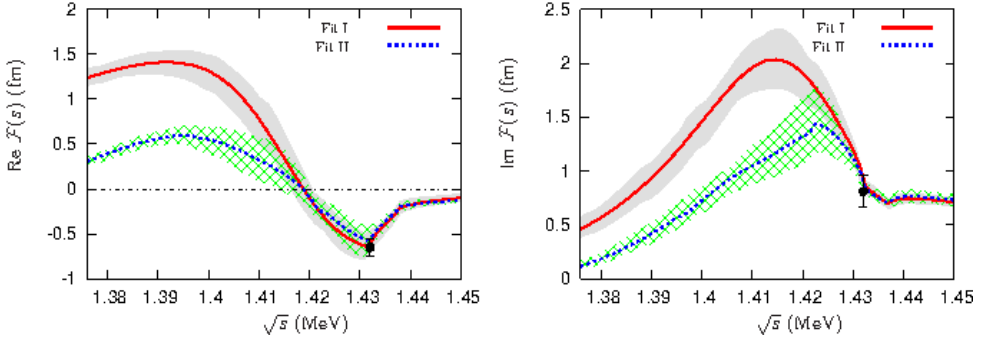} 
\caption{Real (left panel) and imaginary (right panel) parts of the $K^-p$ 
center-of-mass (cm) scattering amplitudes generated in two NLO chiral-model 
fits \cite{GO13}. The $K^-p$ threshold values marked by solid dots follow 
from the SIDDHARTA measurement of kaonic hydrogen $1s$ level shift and 
width \cite{SID11}. Figure adapted from Ref.~\cite{GO13}.} 
\label{fig:oller} 
\end{figure}

The $\bar K N$ interaction near and below threshold is attractive in models 
that generate dynamically the subthreshold $s$-wave resonance $\Lambda(1405)$, 
providing sound motivation to search for $K^-$ bound states in nuclei 
\cite{hyodo13}. Subthreshold $K^-N$ scattering amplitudes are needed for 
calculating such states, even in kaonic atoms for which the kaon energy is 
essentially at threshold \cite{wycech71,BT72,rook75}. However, subthreshold 
$\bar K N$ scattering amplitudes are highly model dependent, as demonstrated 
in Fig.~\ref{fig:oller} showing that two distinct NLO chiral-model fits 
to scattering and reaction data above and at threshold could generate 
$K^-p$ scattering amplitudes differing substantially from each other in 
the subthreshold region. Fit-I amplitude, nevertheless, is quite similar 
to the NLO amplitudes derived by Ikeda, Hyodo and Weise (IHW) \cite{IHW11} 
and by Ciepl\'{y} and Smejkal (CS) \cite{CS12}, both of which were used 
in our recent calculations. 

The present review follows Refs.~\cite{Gal13,Gal14}, highlighting recent 
progress made by the Prague-Jerusalem Collaboration towards incorporating 
medium modifications, particularly those implied by the energy dependence 
of $K^-N$ scattering amplitudes \cite{CFGGM11,CFGK11,BGL12,FG12,GM12,FG13}. 
In-medium $\bar KN$ scattering amplitudes are discussed in Sect.~\ref{sec2}, 
focusing on the connection between their (subthreshold) energy and density 
dependencies. The use of such in-medium $K^-N$ scattering amplitudes 
in kaonic-atom calculations and fits is discussed in Sect.~\ref{sec3}. 
Related applications to kaonic bound-state calculations are discussed 
in Sect.~\ref{sec4} for few-body systems, and in Sect.~\ref{sec5} for 
many-body systems. A brief summary and outlook in Sect.~\ref{sec6} 
concludes this presentation.

\section{In-medium amplitudes and energy versus density dependence}  
\label{sec2} 

Here we follow Ciepl\'{y} and Smejkal \cite{CS12} who introduced meson-baryon 
coupled-channel energy-dependent separable $s$-wave interactions matched to 
SU(3) scattering amplitudes in up to next-to-leading order (NLO) of the chiral 
expansion for strangeness $-1$. It is gratifying that the Tomozawa-Weinberg 
LO term provides fair approximation to the corresponding $\bar{K}N$ amplitudes 
\cite{IHW11,CS12}. Solving the in-medium coupled-channel Lippmann-Schwinger 
equations $F=V+VGF$ with these potential kernels leads to a separable form of 
in-medium scattering amplitudes $F_{ij}$, given in the two-body cm system by 
\begin{equation}
F_{ij}(k,k';\sqrt{s},\rho)=g_{i}(k^{2}) \: f_{ij}(\sqrt{s},\rho) \: 
g_{j}(k'^{2}) \; ,
\label{eq:Fsep}
\end{equation}
with momentum-space form factors $g_{j}(k^{2})$, where $j$ runs over channels, 
and in-medium reduced amplitudes $f_{ij}(\sqrt{s},\rho)$ expressed as 
\begin{equation}
f_{ij}(\sqrt{s}, \rho)=\left[ (1 - v(\sqrt{s}) \cdot G(\sqrt{s}, \rho))^{-1}
\cdot v(\sqrt{s}) \right]_{ij} \; .
\label{eq:fij}
\end{equation}
Here, $G$ is a channel-diagonal Green's function in the nuclear medium: 
\begin{equation}
G_{n}(\sqrt{s},\rho) = -4\pi \: \int_{\Omega_{n}(\rho)}
\frac{d^{3}p}{(2\pi)^{3}}\frac{g_{n}^{2}(p^{2})}
{k_{n}^{2}-p^{2} -\Pi^{(n)}(\sqrt{s},\rho) +{\rm i}0} \; ,
\label{eq:Grho}
\end{equation} 
where the integration on intermediate meson-baryon momenta is limited to 
a region $\Omega_{n}(\rho)$ ensuring that the intermediate nucleon energy 
is above the Fermi level in channels $n$ involving nucleons. The self-energy 
$\Pi^{(n)}(\sqrt{s},\rho)$ stands for the sum of hadron self-energies 
in channel $n$. Of particular interest is the meson ($h$) self-energy 
$\Pi^{(hN)}_h=(E_N\,/\sqrt{s})\,\Pi_h$ in the diagonal $n\equiv (hN)$ channel, 
where the lab self-energy $\Pi_h$ is given by   
\begin{equation} 
\Pi_h(\sqrt{s},\rho)\equiv 2\omega_hV_h=-\frac{\sqrt{s}}{E_N}\,4\pi
F_{hN}(\sqrt{s},\rho)\rho \; ,
\label{eq:Pi} 
\end{equation} 
depending implicitly on $\omega_h=m_h-B_h$ and on the off-shell two-body 
momenta $k,k'$. This self-energy, once evaluated {\it self-consistently} 
while converting its $\sqrt{s}$~~dependence into a full density dependence, 
serves as input to the Klein-Gordon bound-state equation 
\begin{equation} 
[\:\nabla^2+{\tilde\omega}_h^2-m_h^2-\Pi_h(\omega_h,\rho)\:]\:\psi=0  
\; , 
\label{eq:KG} 
\end{equation} 
in which ${\tilde\omega}_h=\omega_h-{\rm i}\Gamma_h/2$, with $B_h$ and 
$\Gamma_h$ the binding energy and the width of the meson-nuclear bound state, 
respectively. 

\begin{figure}[htb] 
\begin{center} 
\includegraphics[width=0.48\textwidth,height=4.6cm]{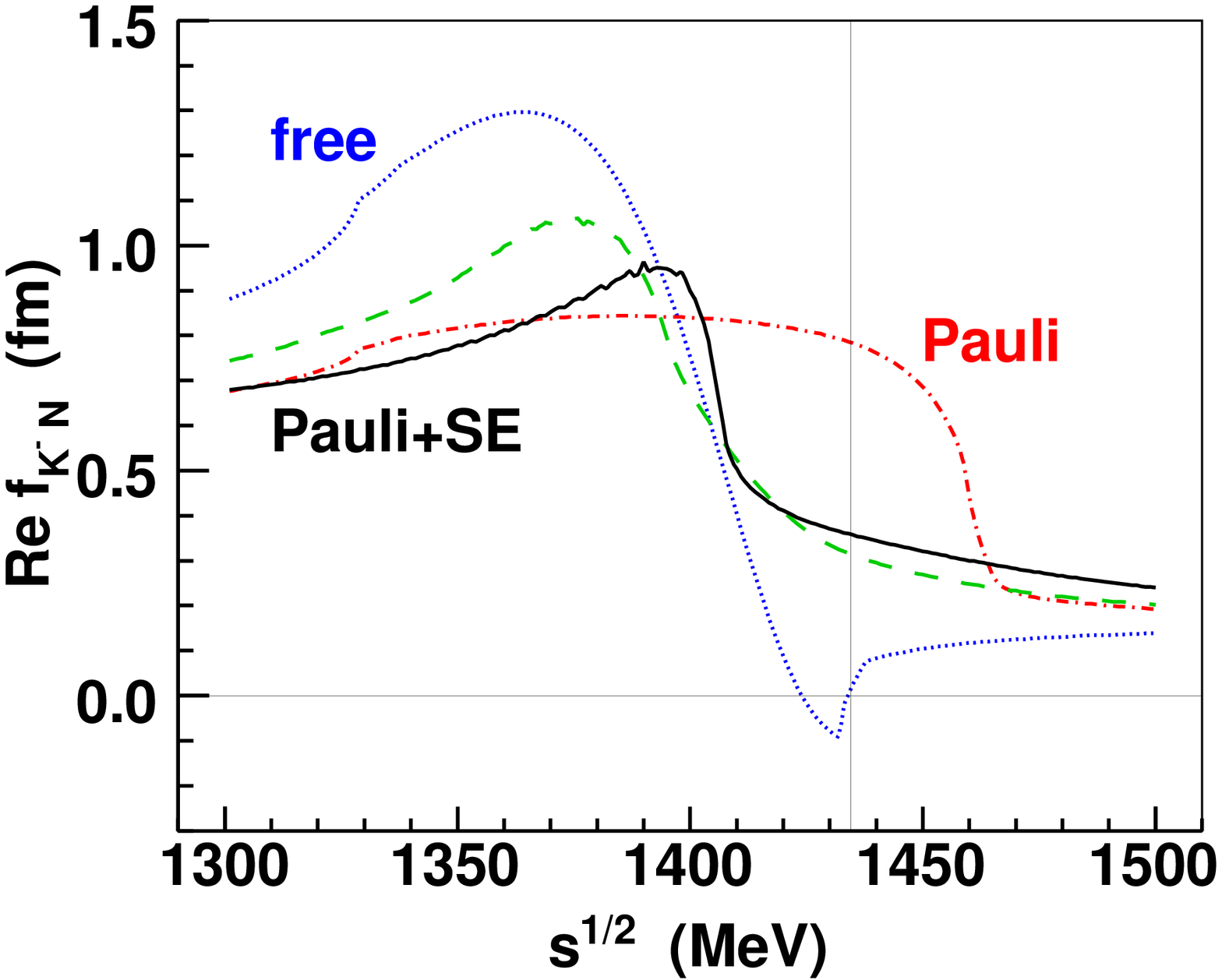} 
\includegraphics[width=0.48\textwidth,height=4.6cm]{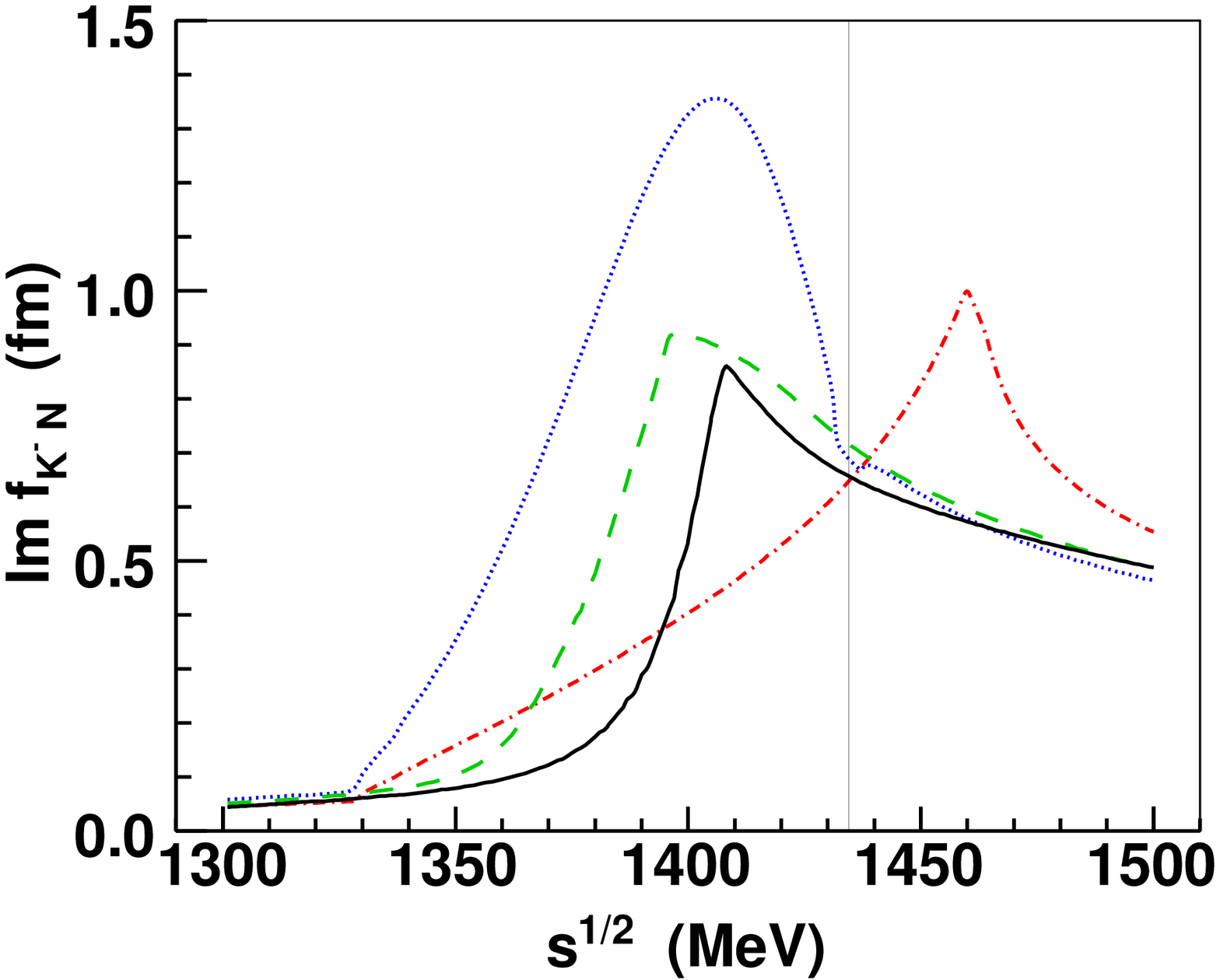} 
\caption{Near-threshold energy dependence of $K^-N$ cm reduced 
scattering amplitudes (left: real, right: imaginary) in model NLO30 of 
Ref.~\cite{CS12} for free-space (dotted) and Pauli-blocked amplitudes at 
$\rho=\rho_0$ with (solid) and without (dot-dashed) meson and baryon 
self-energies (SE). The dashed curves show Pauli-blocked amplitudes with 
SE at $\rho=0.5\rho_0$. The $K^-N$ threshold is marked by a thin vertical 
line.} 
\label{fig:NLO30} 
\end{center} 
\end{figure} 

The resulting in-medium $K^-N$ isoscalar amplitudes above and below threshold 
are shown in Fig.~\ref{fig:NLO30}. The real part of the amplitude is strongly 
attractive at subthreshold energies that according to the discussion below are 
relevant to $K^-$ atomic and nuclear states. The attraction as well as the 
imaginary-part absorptivity get moderately weaker for $\rho\geq 0.5\rho_0$, as 
demonstrated by comparing on the left panel the solid curves ($\rho=\rho_0$) 
with the dashed curves ($\rho=0.5\rho_0$). This implies that $K^-$ bound 
states are very likely to exist, but with rather large widths generated by 
the imaginary-part absorptivity. 

To determine the subthreshold energies for use in in-medium hadron-nucleon 
scattering amplitudes, we consider the downward energy shift $\delta\sqrt{s}
\equiv\sqrt{s}-\sqrt{s_{\rm th}}$ where $\sqrt{s_{\rm th}}\equiv m_{h}+m_N$
and $s=(\sqrt{s_{\rm th}}-B_{h}-B_N)^2-({\vec p}_{h}+{\vec p}_N)^2$, with 
$B_h$ and $B_N$ meson and nucleon binding energies. Since ${\vec p}_{h}+{
\vec p}_N\neq 0$ in the meson-nuclear cm frame (approximately the lab frame), 
the associated negative contribution to $s$ has to be included. To leading 
order in binding energies and kinetic energies with respect to rest masses, 
and specializing to $\bar K$ mesons (denoted $h=K$), $\delta\sqrt{s}$ is 
expressed as 
\begin{equation} 
\delta\sqrt{s}\approx - B_N - B_K - \xi_N\frac{p_N^2}{2m_N} 
- \xi_K\frac{p_K^2}{2m_K} \; , \;\;\;\;\; 
\xi_{N(K)}\equiv \frac{m_{N(K)}}{m_N+m_K} \; .
\label{eq:approx}
\end{equation} 

\begin{figure}[htb] 
\begin{center} 
\includegraphics[width=0.48\textwidth,height=4.6cm]{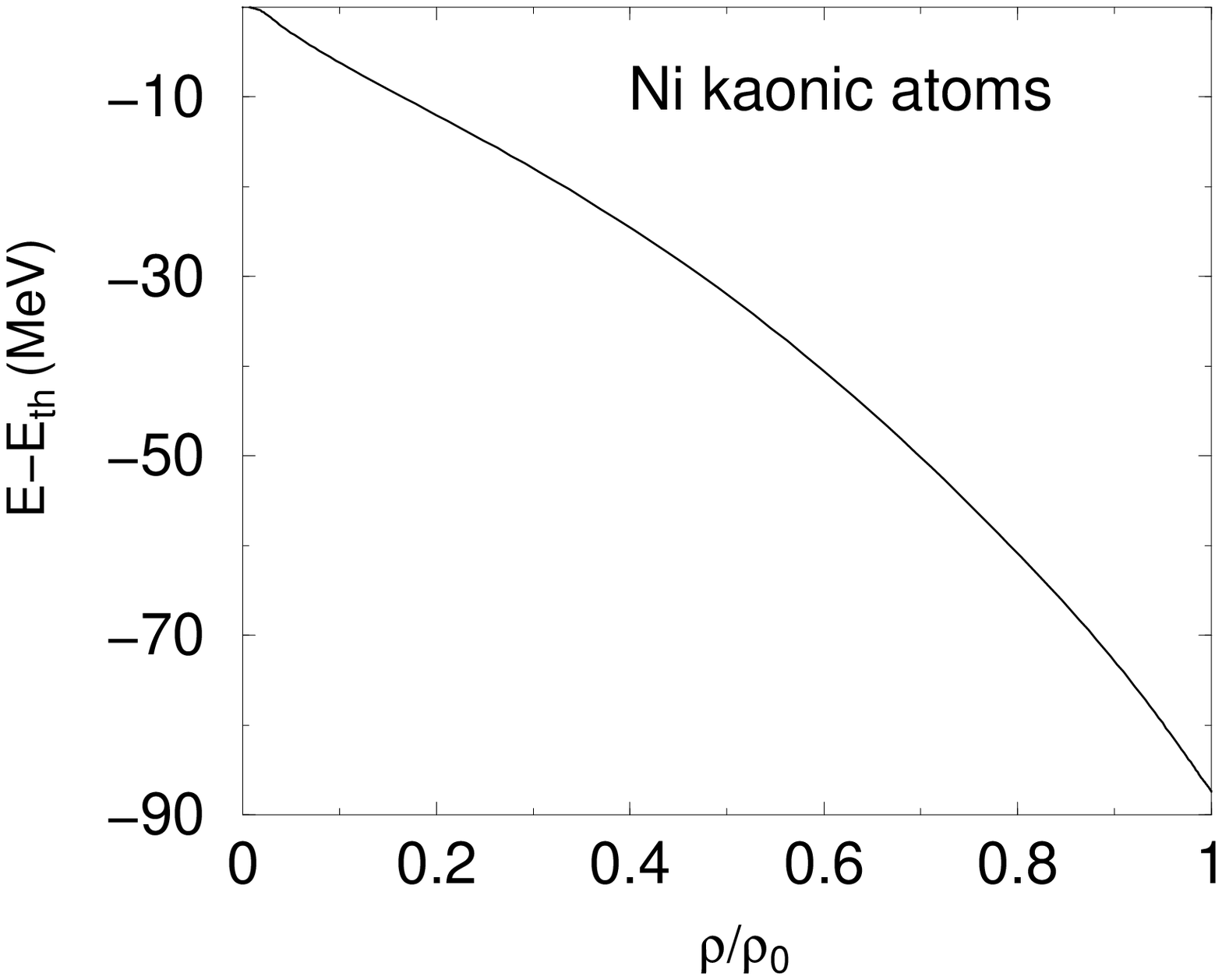} 
\includegraphics[width=0.48\textwidth,height=4.6cm]{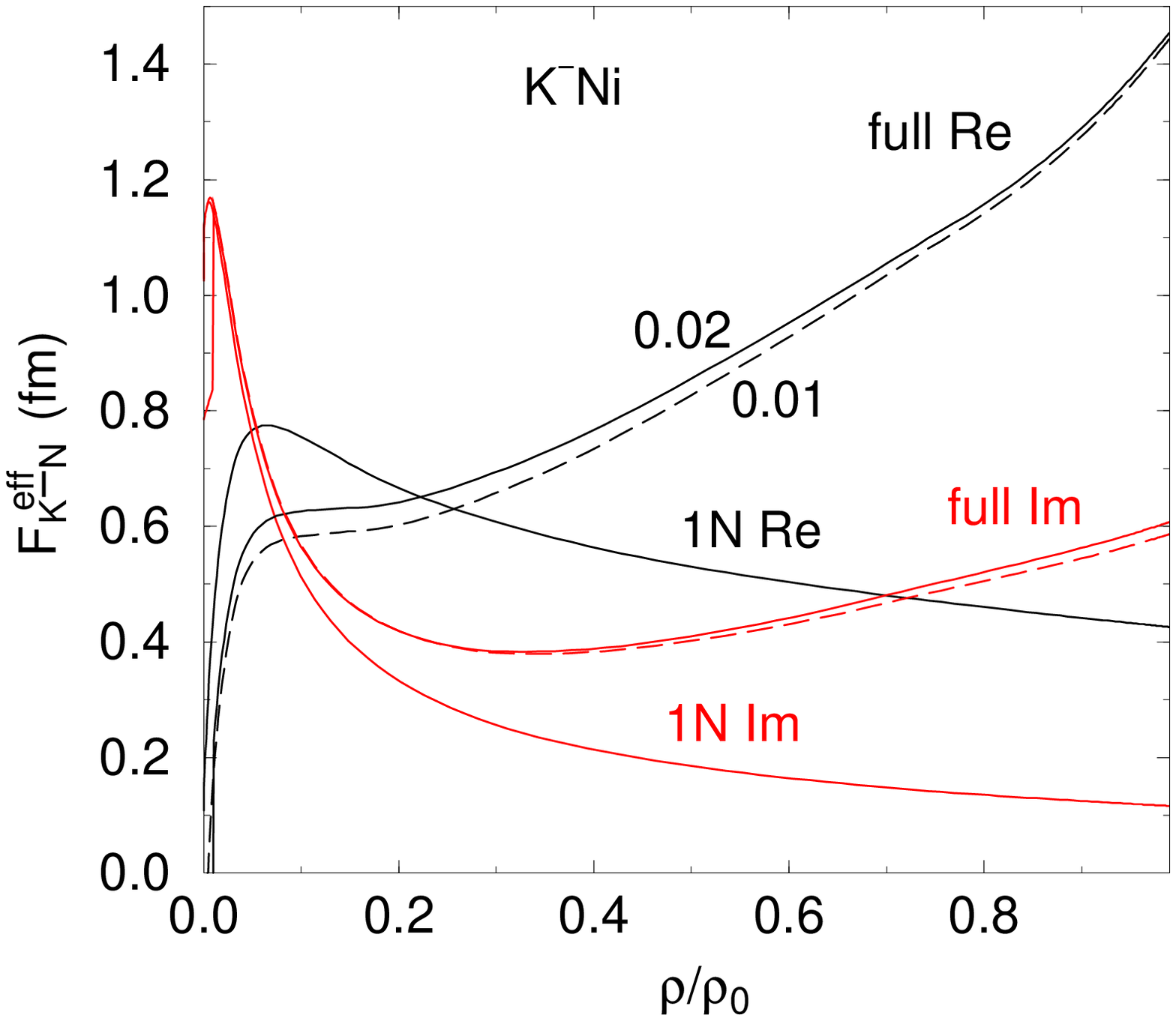} 
\caption{Left: subthreshold energies probed in $K^-$-Ni atom as a function of 
nuclear density, calculated self-consistently within the IHW-based global fit 
to kaonic atoms using Eq.~(\ref{eq:sqrts}). 
Right: Kaonic-atom globally fitted amplitude $F^{\rm eff}_{K^-}(\rho)$, 
marked ``full", and the in-medium IHW-based amplitude $F_{K^-N}(\rho)$ in 
the absence of many-nucleon contributions, marked ``1N", as a function of 
nuclear density in Ni \cite{FG13}. Solid (dashed) curves are for matching 
to free-space amplitudes at 0.02(0.01)$\rho_0$.} 
\label{fig:Evsrho} 
\end{center} 
\end{figure} 

Using the Fermi Gas model for nucleons and the local density approximation for 
$\bar K$, assuming also the minimal substitution (MS) principle \cite{FG14}, 
one obtains 
\begin{equation} 
\delta\sqrt{s}\approx-B_N\frac{\rho}{{\bar\rho}}-\xi_N [B_K\frac{\rho}{\rho_0}
+T_N(\frac{\rho}{\bar\rho})^{2/3}+V_c(\frac{\rho}{\rho_0})^{1/3}]-\xi_K 
\frac{\sqrt{s}}{\omega_K E_N}2\pi\,{\rm Re}\,F_{\bar KN}(\sqrt{s},\rho)\rho\;, 
\label{eq:sqrts} 
\end{equation}
where $V_c$ is the $K^-$ Coulomb potential due to the finite-size nuclear 
charge distribution, $T_N=23.0$ MeV is the average nucleon kinetic energy, 
$B_N\approx 8.5$~MeV is an average nucleon binding energy and $\bar\rho$ 
and $\rho_0$ are the average nuclear density and nuclear-matter density, 
respectively. Expression (\ref{eq:sqrts}) respects the low-density 
limit, $\delta\sqrt{s}\to 0$ upon $\rho\to 0$. For attractive scattering 
amplitudes, the last term of (\ref{eq:sqrts}), in particular, provides 
substantial downward energy shift overlooked by many previous calculations 
that assumed ${\vec p}_K=0$ which is inappropriate for {\it finite} nuclei. 
Since $\sqrt{s}$~~depends through Eq.~(\ref{eq:sqrts}) on ${\rm Re}~F_{\bar KN}
(\sqrt{s},\rho)$ which by itself depends on $\sqrt{s}$, it is clear 
that for a given value of $B_K$, $F_{\bar KN}(\sqrt{s},\rho)$ has to be 
determined {\it self-consistently} by iterating Eq.~(\ref{eq:sqrts}). 
This is done at each radial point where $\rho$ is given, and for each $B_K$ 
value during the calculation of bound states. The emerging correlation between 
the downward energy shift $\delta$$\sqrt{s}$~~and the density $\rho$ renders 
$F_{\bar KN}(\sqrt{s},\rho)$ into a state-dependent function of the density 
$\rho$ alone, denoted for brevity by $F_{\bar KN}(\rho)$. This correlation 
is shown on the left side of Fig.~\ref{fig:Evsrho} for kaonic atoms, where 
$B_{K^-}\approx 0$. The figure demonstrates appreciable energy shifts below 
threshold in kaonic atoms, although these are somewhat smaller 
(by $\sim$10~MeV in Ni) than the shifts evaluated in \cite{FG13} without 
incorporating MS.

\section{$K^-$ interactions in kaonic atoms} 
\label{sec3} 

The most recent kaonic-atom calculations are due to Friedman and Gal in 
Ref.~\cite{FG12}, using in-medium $K^-N$ scattering amplitudes generated from 
the Ciepl\'{y}-Smejkal (CS) NLO30 model as described in the previous section, 
and in Ref.~\cite{FG13} using Pauli blocked $K^-N$ scattering amplitudes 
generated from the Ikeda-Hyodo-Weise (IHW) NLO work. The CS \cite{CS12} and 
IHW \cite{IHW11} free-space amplitudes $F_{K^-N}(\sqrt{s})$ agree 
semi-quantitatively with each other. The kaonic-atom fit in Ref.~\cite{FG13} 
considers in addition to the in-medium IHW-based one-nucleon (1N) amplitude 
$F_{K^-N}(\sqrt{s},\rho)$ input also many-nucleon absorptive and dispersive 
contributions, represented by energy-independent phenomenological amplitude 
$F^{\rm many}_{K^-N}(\rho)$ with prescribed density dependence form that 
includes several fit parameters. The assumption of energy independence 
is motivated by observing that $K^-$ absorption on two nucleons, which is 
expected to dominate $F^{\rm many}_{K^-N}$, releases energy $\sim$$m_{K^-}$ 
considerably larger than the subthreshold energies of less than 100 MeV 
encountered in kaonic-atom calculations. The self-energy input $\Pi_{K^-}$ 
to the KG equation (\ref{eq:KG}) is now constructed from an {\it effective} 
$K^-N$ scattering amplitude $F^{\rm eff}_{K^-N}=F^{\rm one}_{K^-N}+F^{
\rm many}_{K^-N}$ which is iterated through the self-consistency expression 
(\ref{eq:sqrts}). This introduces coupling between the many-nucleon fitted 
amplitude $F^{\rm many}_{K^-N}$ and the converged one-nucleon amplitude 
$F^{\rm one}_{K^-N}$ that evolves from the 1N input amplitude 
$F_{K^-N}(\rho)$: $F^{\rm one}_{K^-N}(\rho)\to F_{K^-N}(\rho)$ upon 
$F^{\rm many}_{K^-N}\to 0$. 

The full effective amplitude $F^{\rm eff}_{K^-N}(\rho)$ resulting from 
the global kaonic-atom fit in \cite{FG13} is shown in the right panel of 
Fig.~\ref{fig:Evsrho} marked ``full", along with the in-medium IHW-based 
amplitude $F_{K^-N}(\rho)$ marked ``1N". The figure makes it clear that for 
densities exceeding $\sim$0.5$\rho_0$ the full effective amplitude departs 
appreciably from the in-medium IHW-based amplitude, which in the case of the 
imaginary part amounts to doubling the 1N absorptivity of in-medium $K^-$ 
mesons. For more details, see \cite{FG13}. 

\begin{figure}[htb] 
\begin{center} 
\includegraphics[width=0.48\textwidth,height=7.0cm]{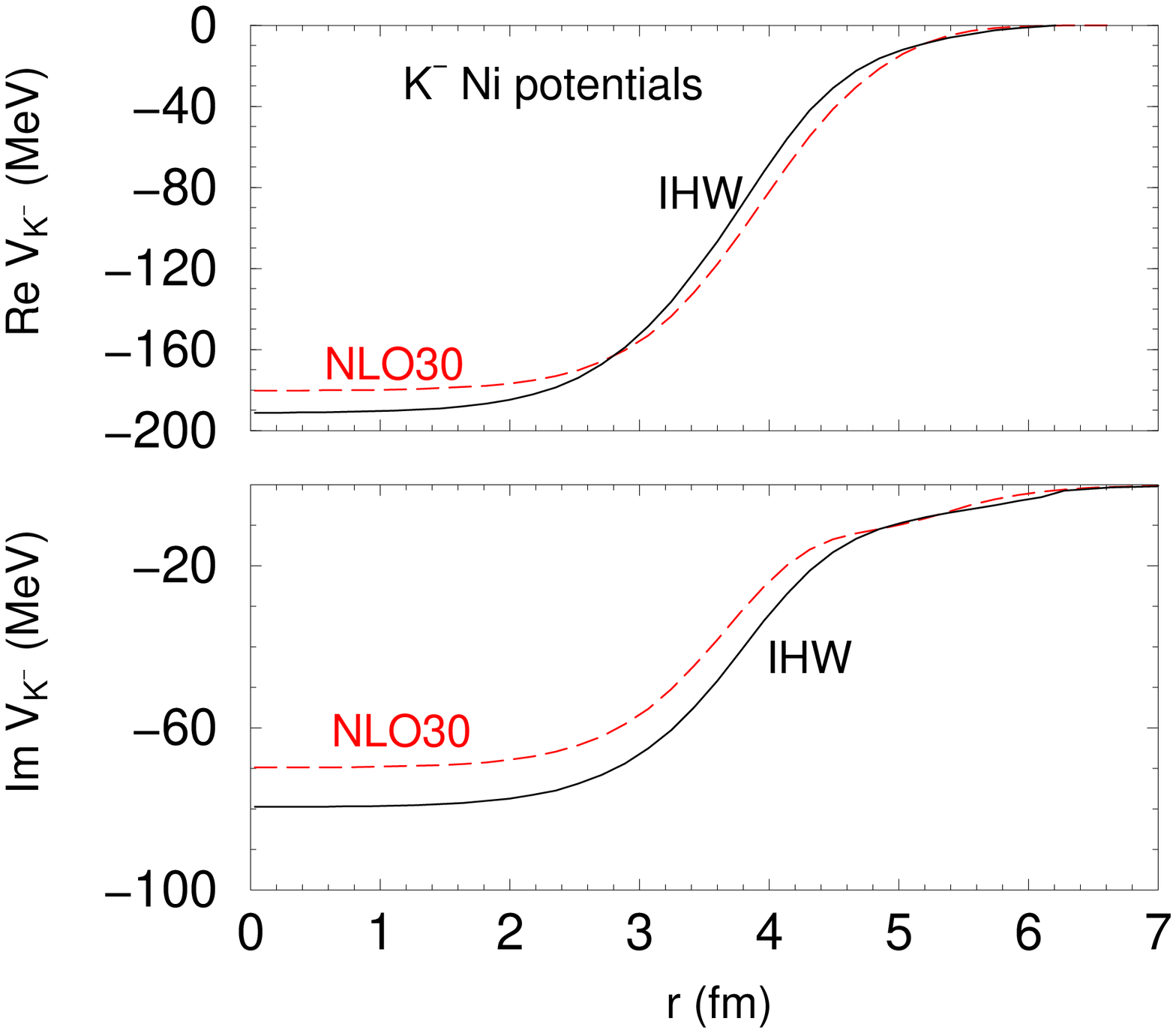} 
\includegraphics[width=0.48\textwidth,height=7.0cm]{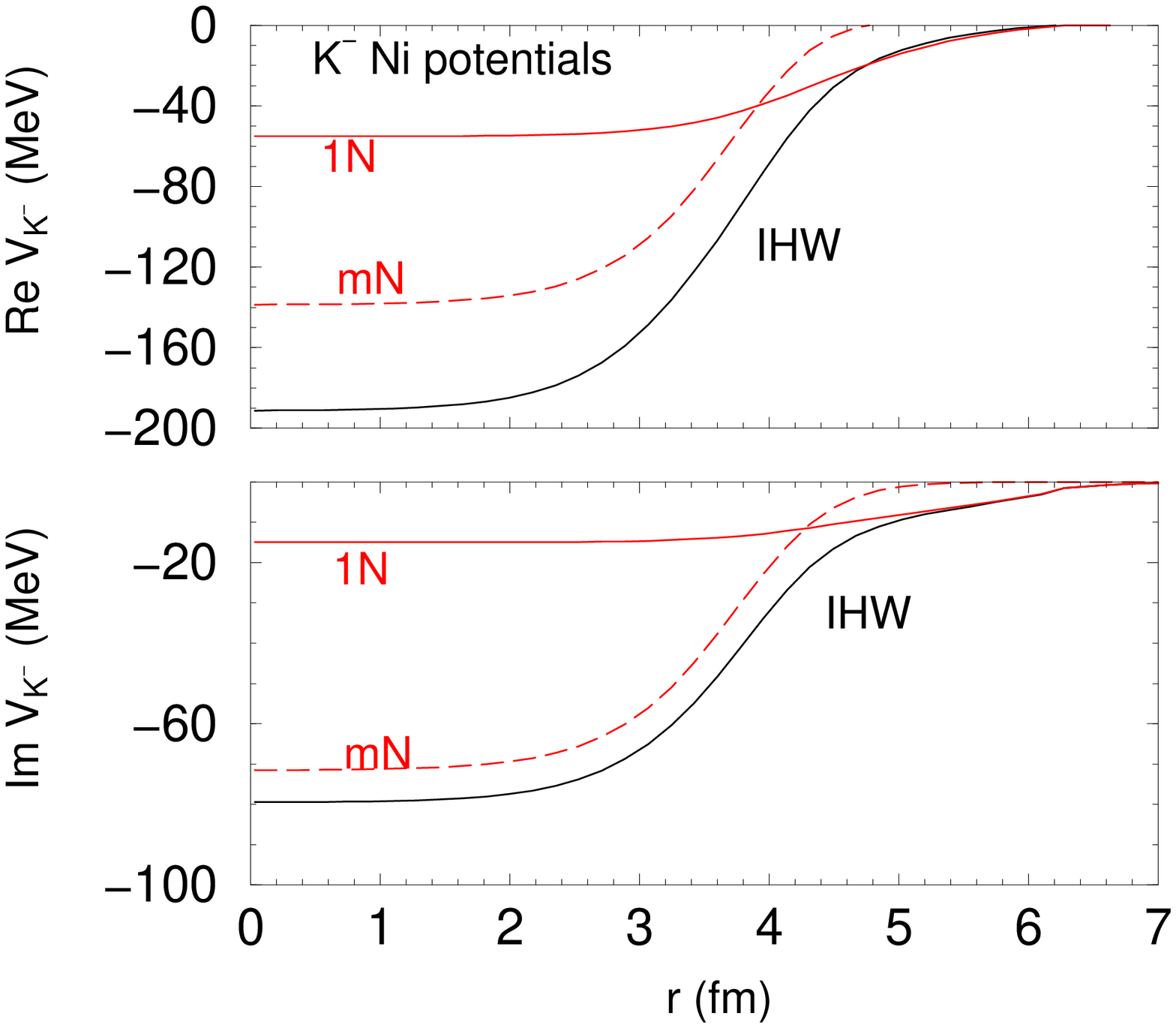} 
\caption{Left: Self-consistent $K^-$ nuclear potentials $V_{K^-}$ for $K^-$ 
atoms of Ni derived from global fits based on in-medium IHW amplitudes 
\cite{FG13}, with the corresponding 1N and many-nucleon (mN) components 
on the right panel. The dashed curves in the left panel are derived from 
in-medium NLO30 amplitudes \cite{FG12}.} 
\label{fig:Kpot} 
\end{center} 
\end{figure} 

The $K^-$ nuclear attraction and absorptivity deduced from global kaonic-atom 
fits are sizable at central nuclear densities. This is demonstrated for Ni in 
Fig.~\ref{fig:Kpot} by the real and imaginary parts of the potential $V_{K^-}$ 
defined by (\ref{eq:Pi}). Although the potential depths might reflect merely a 
smooth extrapolation provided by the input components of the $K^-N$ amplitude, 
the potential at 0.5$\rho_0$ and perhaps up to 0.9$\rho_0$ is reliably 
determined in kaonic-atom fits \cite{BF07}. It is reassuring that both 
IHW-based and NLO30-based fits agree with each other semi-quantitatively 
as shown on the left panel. The right panel of Fig.~\ref{fig:Kpot} shows 
a non-additive splitting of the fitted $K^-$-nuclear potential into a 1N 
in-medium component, obtained on the assumption that there is no many-nucleon 
(mN) component present, and a fitted mN component. The composition of the 
imaginary part of the potential is of particular interest, indicating that 
the mN component which is sizable in the nuclear interior becomes negligible 
about half a fermi outside of the half-density radius. This has implications 
for choosing optimally kaonic-atom candidates where widths of two atomic 
levels can be measured so as to substantiate the 1N vs mN pattern observed 
in global fits \cite{FO13}.

\section{Few-body kaonic quasibound states}
\label{sec4}

\begin{figure}[htb] 
\begin{center} 
\includegraphics[width=0.48\textwidth,height=5cm]{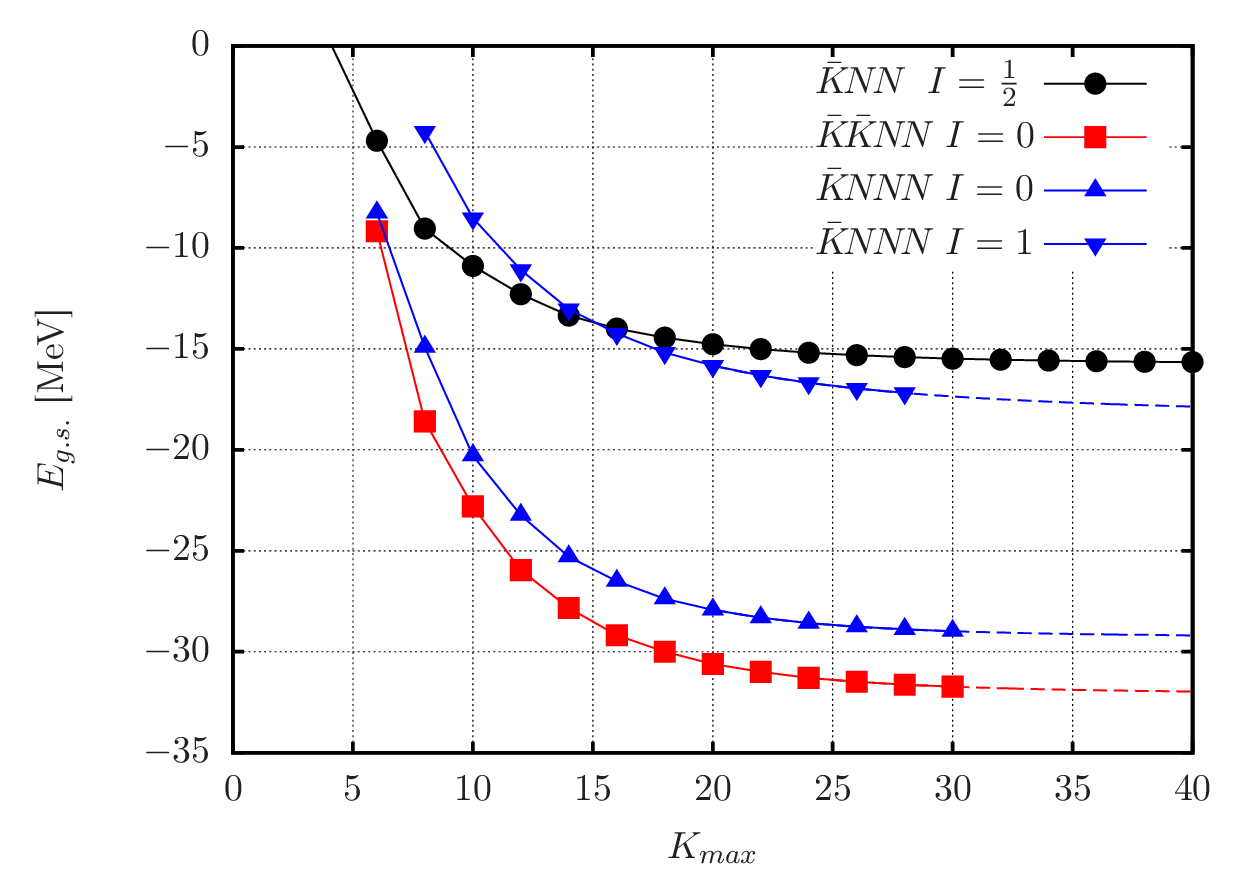} 
\includegraphics[width=0.48\textwidth,height=5cm]{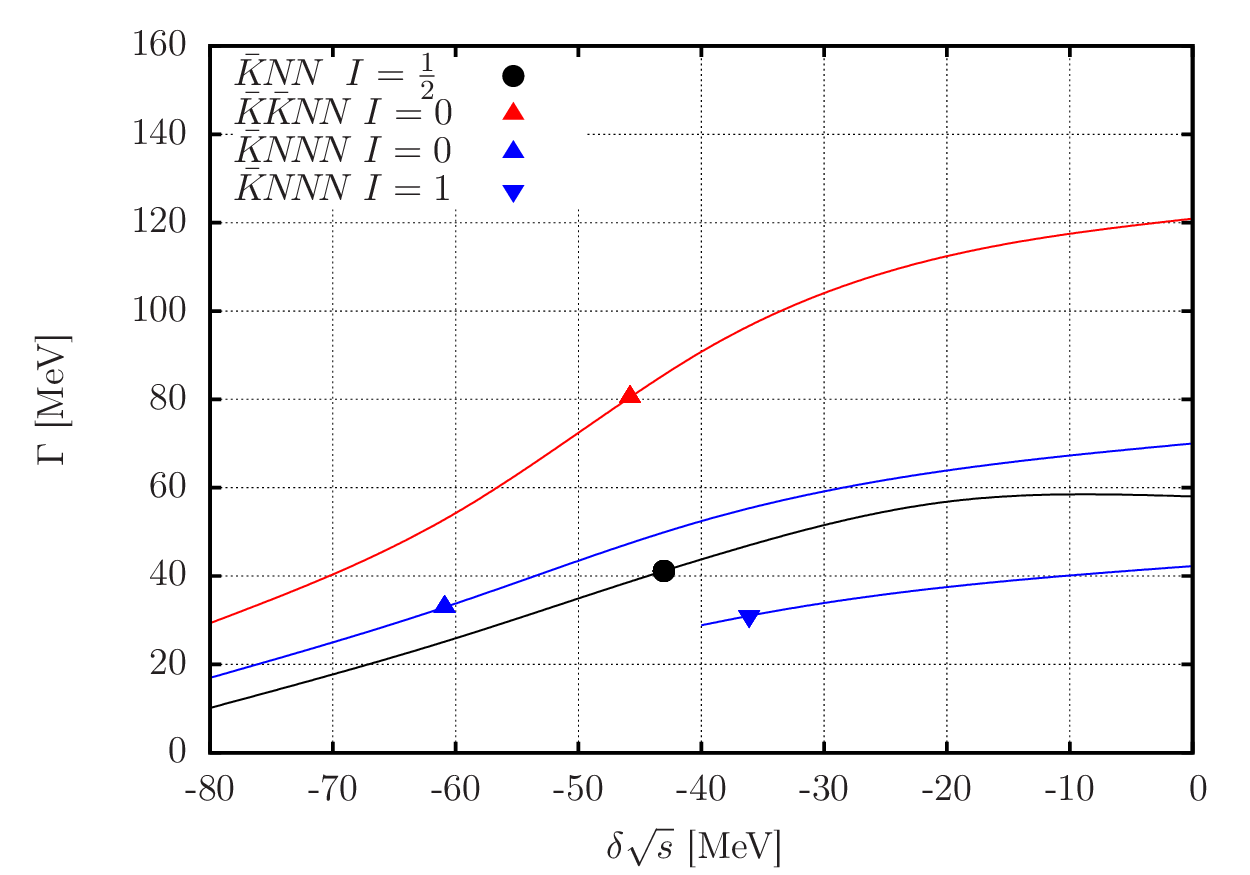} 
\caption{Left: binding energies of $\bar K$ and $\bar K\bar K$ few-body 
quasibound states from Ref.~\cite{BGL12} calculated self-consistently and 
plotted here as a function of the maximal total grand angular momentum 
$K_{\rm max}$ in the hyperspherical basis. Right: $\bar K N \to \pi Y$ 
widths (\ref{eq:Gamma}) plotted as a function of 
$\delta\sqrt{s}$~~(\ref{eq:sqrt{s}}).}
\label{fig:BGL} 
\end{center} 
\end{figure}

For $K^-$-nuclear three- and four-body calculations, a variant of the downward 
energy shift Eq.~(\ref{eq:sqrts}) derived for many-body calculations was 
formulated by Barnea, Gal and Liverts \cite{BGL12}: 
\begin{equation} 
\delta\sqrt{s} = -\frac{B}{A}-\frac{A-1}{A}B_K-
\xi_{N}\frac{A-1}{A}\langle T_{N:N} \rangle -\xi_{K}\left ( \frac{A-1}{A} 
\right )^2 \langle T_K \rangle \; ,
\label{eq:sqrt{s}} 
\end{equation} 
with $A$ the baryonic number, $B$ the total binding energy of the system, 
$B_K=-E_K$, $T_K$ the kaon kinetic energy operator in the total cm frame 
and $T_{N:N}$ the pairwise $NN$ kinetic energy operator in the $NN$ pair 
cm system. Note that $\delta$$\sqrt{s}$~~is negative-definite by expression 
(\ref{eq:sqrt{s}}) which provides a self-consistency cycle upon requiring 
that $\sqrt{s}$~~derived through Eq.~(\ref{eq:sqrt{s}}) from the solution of 
the Schroedinger equation agrees with the value of $\sqrt{s}$~~used for the 
input $V_{\bar K N}(\sqrt{s})$. Converged total binding energies calculated 
variationally in the hyperspherical basis are shown on the left panel 
of Fig.~\ref{fig:BGL} for three- and four-body kaonic bound states. The 
corresponding $\bar K N\to\pi Y$ widths, calculated using the approximation 
\begin{equation} 
\frac{\Gamma}{2}\approx\langle \,\Psi_{\rm g.s.} |
-{\rm Im}\,{\cal V}_{\bar KN}\, | \, \Psi_{\rm g.s.} \, \rangle \;, 
\label{eq:Gamma} 
\end{equation} 
where ${\cal V}_{\bar KN}$ consists of all pairwise $\bar KN$ interactions, 
are plotted on the right panel as a function of $\delta\sqrt{s}$, with 
self-consistent values marked on each one of the $\Gamma$-vs-$\delta\sqrt{s}$ 
~curves. Eq.~(\ref{eq:Gamma}) provides a good approximation owing to 
$|{\rm Im}\,{\cal V}_{\bar KN}|\ll|{\rm Re}\,{\cal V}_{\bar KN}|$ \cite{HW08}. 
Expressions similar to (\ref{eq:sqrt{s}}) and (\ref{eq:Gamma}) were used 
in ${\bar K}{\bar K}NN$ calculations. For details of the chirally-based 
energy-dependent $\bar K N$ interaction input and the actual calculations 
of these few-body kaonic clusters, see Ref.~\cite{BGL12}.  

With $\bar K N$ interactions that become weaker upon going subthreshold 
\cite{HW08}, the self-consistently calculated binding energies (widths) come 
out typically 10 (10--40) MeV lower than for threshold input interactions 
$V_{\bar KN}(\sqrt{s_{\rm th}})$, as exhibited in the right panel by comparing 
the marked self-consistent values of $\Gamma$ to the threshold values 
$\Gamma(\delta\sqrt{s}=0)$, and in agreement with the recent Faddeev and 
Faddeev-Yakubovsky calculations by Maeda et al. \cite{maeda13} who used 
energy-independent interactions. In particular, the $I=1/2$ $\bar KNN$ g.s. 
known as `$K^-pp$' is weakly bound, lying just 4.3 MeV below the lowest 
threshold of $N$+$(\bar KN)_{I=0}$ at $-$11.4~MeV. The latter value differs 
substantially from the $-$27~MeV assigned traditionally to the $\Lambda(1405)$ 
resonance and used in non-chiral calculations. Smaller differences occur in 
chiral models, where two $I=0$ poles appear, the upper of which is often 
identified with a $(\bar KN)_{I=0}$ quasibound state. For example, suppressing 
widths, the difference between the upper-pole positions in the IHW chiral 
model \cite{IHW11} (which is close to the one used in \cite{BGL12}) and in 
the recent R\'{e}vai-Shevchenko (RS) chirally motivated model \cite{RS14} 
is $\Delta E(\bar KN)_{I=0}=7$~MeV. A rough estimate of the model dependence 
expected for $K^-pp$ is $\Delta E(K^-pp)\approx 1.5\, \Delta E(\bar KN)_{I=0}
=10.5$~MeV. Thus, a single-channel $\bar KNN$ three-body calculation done 
{\it \`{a} la} Barnea et al. \cite{BGL12} using the RS model is expected to 
give $E(K^-pp)\approx -26$~MeV. This has to be compared with $E=-32$~MeV 
obtained in the RS $\bar KNN-\pi\Sigma N$ coupled-channel Faddeev 
calculation \cite{RS14}. The 6~MeV missing in this crude estimate is of 
the order of magnitude expected \cite{ikeda09} from upgrading a $\bar KNN$ 
single-channel calculation to a $\bar KNN-\pi\Sigma N$ coupled-channel 
calculation. In this respect we reject the unfounded criticism made by 
RS of the self-consistency procedure application in Ref.~\cite{BGL12}. 

The widths exhibited in the figure, of order 40 MeV for single-$\bar K$ 
clusters and twice that for double-$\bar K$ clusters, are due to $\bar KN\to 
\pi Y$. Additional $\bar K NN\to YN$ contributions of up to $\sim$10~MeV in 
$K^-pp$ \cite{DHW08} (see, however, the estimate $\Gamma_{\bar K NN\to YN}
\sim 30$~MeV made in Ref.~\cite{oset13}) and $\sim$20~MeV in the four-body 
systems \cite{BGL12} are foreseen. Altogether, widths of order 50 MeV or 
higher are anticipated for few-body kaonic quasibound states.

\section{Many-body kaonic quasibound states} 
\label{sec5} 

In-medium $\bar KN$ scattering amplitudes derived from the chirally motivated 
NLO30 model \cite{CS12} were employed by Gazda and Mare\v{s} \cite{GM12} to 
evaluate self-consistently $K^-$ quasibound states across the periodic table, 
using static RMF nuclear-core densities. Calculated $K^-$ binding energies 
and widths in Ca are listed in Table~\ref{tab:GM12} for several choices of 
input interactions.  

\begin{table}[htb!]
\centering
\caption{Self-consistently calculated binding energies $B_K$ and widths 
$\Gamma_K$ (in MeV) of $K^-$ quasibound states in Ca, using a static RMF Ca 
density and NLO30 in-medium $K^-N$ subthreshold amplitudes, see text.} 
\begin{tabular}{|l|cc|cc|cc|}
\hline 
 & \multicolumn{2}{|c|}{NLO30} & \multicolumn{2}{|c|}{+ $p$ wave} & 
\multicolumn{2}{|c|}{+ $2N$ abs.} \\ 
 & $B_K$ & $\Gamma_K$ & $B_K$ & $\Gamma_K$ & $B_K$ & $\Gamma_K$ \\
\hline 
$1s_K$ & 70.5 & 14.9 & 73.0 & 14.8 & 68.9 & 58.9 \\ 
$1p_K$ & 50.6 & 18.0 & 53.1 & 17.9 & 49.2 & 53.6 \\ 
$1d_K$ & 28.8 & 30.3 & 32.1 & 29.3 & 27.7 & 59.7 \\ 
$2s_K$ & 23.9 & 33.8 & 26.3 & 34.2 & 21.6 & 67.1 \\ 
\hline 
\end{tabular} 
\label{tab:GM12} 
\end{table} 

In addition to $B_K$ and $\Gamma_K$ values for NLO30 in-medium $s$-wave $K^-N$ 
interactions, we listed in Table~\ref{tab:GM12} values derived (i) by adding 
a $\Sigma(1385)$-motivated $p$-wave $K^-N$ interaction from Ref.~\cite{WH08}, 
thereby increasing $B_K$ marginally by a few MeV and modifying $\Gamma_K$ 
by less than 1 MeV, or (ii) by adding a two-nucleon ($2N$) $K^-NN$$\to$$YN$ 
absorption term estimated from fitting to kaonic atoms, resulting in 
$\lesssim$2~MeV decrease of $B_K$ but substantially increasing the width to 
$\Gamma_K\sim (50-70)$~MeV. Given these large widths, it is unlikely that 
distinct quasibound states can be uniquely resolved, except perhaps in very 
light $K^-$ nuclei.  

The hierarchy of widths listed in Table~\ref{tab:GM12} is also worth noting. 
With energy-independent potentials one expects maximal widths for the lowest, 
most localized 1$s_K$ states, and gradualy decreased widths in excited states 
which are less localized within the nucleus. The reverse is observed here, 
in particular upon excluding $2N$ absorption. This is a corollary of requiring 
self consistency: the more excited a $K^-$ quasibound state is, the lower 
nuclear density it feels, and a smaller downward shift into subthreshold 
energies it probes via the $\sqrt{s(\rho)}$ dependence. 
Since Im$\,f_{K^-N}(\rho)$ decreases strongly upon going below threshold, 
see Fig.~\ref{fig:NLO30}, its contribution to the calculated width gets 
larger, the higher the excited quasibound-state energy is.

\section{Summary and outlook} 
\label{sec6} 

In this overview of $\bar K$-nuclear bound-state calculations we 
have focused on the role played by the underlying meson-baryon subthreshold 
dynamics. It was shown how the energy dependence of the meson-baryon 
in-medium scattering amplitudes is converted into density dependence 
of the meson self-energies, or equivalently of meson-nucleus optical 
potentials. Based on global fits of $K^-$-atom data we argued that the 
in-medium chiral model input has to be supplemented by appreciable 
many-nucleon dispersive and absorptive potential contributions which imply 
uniformly large widths of order 50 MeV and more for $\bar K$-nuclear bound 
states. The experimental thrust at present and in the near future hinges on 
$K^-pp$ searches, as reviewed by Nagae in this conference \cite{nagae14}.

\begin{acknowledgement}

\noindent This work was supported by the GACR Grant No. 203/12/2126, 
as well as by the EU initiative FP7, HadronPhysics3, under the SPHERE 
and LEANNIS cooperation programs. 

\end{acknowledgement}

\end{document}